\documentclass[superscriptaddress,amsmath,amssymb,
aps, prstab, twocolumn, groupedaddress]{revtex4-2} 
\usepackage{amsmath}
\usepackage{natbib}
\usepackage{float}
\usepackage{amssymb}
\usepackage{physics}
\usepackage{dcolumn}
\usepackage{graphicx}
\usepackage{longtable}
\usepackage{bm}
\usepackage{color}
\usepackage{epsfig}
\usepackage{xr-hyper}
\usepackage{hyperref}

\usepackage{enumerate}

\begin{document}

\title{Finite Spinon Density-of-States in Triangular-Lattice Delafossite TlYbSe$_2$}

\author{Bishnu P. Belbase$^{1}$}
\thanks{\text{These authors contributed equally to this work}}
\author{Arjun Unnikrishnan$^{1,2,\ast}$}
\email{arjunu@iisc.ac.in}
\author{Shi Feng$^{3,4}$}
\author{Eun Sang Choi$^{5}$}
\author{Johannes Knolle$^{3,4,6}$}
\author{Arnab Banerjee$^{1}$}
\email{arnabb@purdue.edu}

\affiliation{$^1$Department of Physics and Astronomy, Purdue University, West Lafayette, Indiana 47906, USA}
\affiliation{$^2$Solid State and Structural Chemistry Unit (SSCU), Indian Institute of Science, Bengaluru - 560012, India}
\affiliation{$^3$Technical University of Munich, TUM School of Natural Sciences, Physics Department, Garching, Germany}
\affiliation{$^4$Munich Center for Quantum Science and Technology (MCQST), Schellingstr. 4, 80799 M{\"u}nchen, Germany}
\affiliation{$^5$National High Magnetic Field Laboratory, Tallahassee, Florida - 32310, USA}
\affiliation{$^6$Blackett Laboratory, Imperial College London, London SW7 2AZ, United Kingdom}

\date{\today}

\begin{abstract}
\noindent We introduce the rare-earth delafossite compound TlYbSe$_2$  - extending the search for quantum spin liquids in frustrated triangular lattice magnets. While the DC magnetisation suggests magnetic exchange interactions in the order of several Kelvin, the zero-field AC magnetisation and heat capacity measurements reveal no signs of long-range magnetic order down to $20$~mK, indicating a quantum-disordered ground state. We observe a spin glass transition around $\sim30$~mK at zero field, arguably originating from a small fraction of free spins--with an associated entropy of $<3\%$ of the total $R\ln 2$, which is suppressed by an applied field of $\sim0.02$~T. A broad anomaly in the heat capacity measurements between $2-5$~K is indicative of short-range spin correlations. Below $350$~mK, we observe a robust linear temperature dependence of the heat capacity, accompanied by the complete absence of long-range order at low fields. We propose that a phenomenological theory, based on the interplay between spinons and thermally excited gauge flux excitations, can account for the linear temperature dependence of the heat capacity, and could be widely applicable to similar critical quantum spin liquid candidate materials. The results establish the low-temperature, low-field regime of TlYbSe$_2$ as a prime candidate for field-tunable triangular quantum spin liquid behavior and highlight the importance of thermally excited gauge field excitations.

\end{abstract}

\maketitle

{\it Introduction.--} Quantum spin liquids (QSLs) are exotic states of matter sought after in low-dimensional frustrated magnets with potential applications ranging from quantum memory to topologically protected qubits~\cite{Dennis2002topological,balents2010spin,savary2016quantum,knolle2019field,zhou2017quantum,hirobe2017one,broholm2020quantum,li2015rare,bordelon2020spin,arh2022ising,shimizu2003spin,helton2007spin,han2012fractionalized,feng2017gapped,banerjee2016proximate,meng2010quantum}. The two-dimensional (2D) triangular lattice stands out as Anderson's original proposal for a simple QSL platform~\cite {anderson1973resonating,kanoda2011mott,balents2010spin,iqbal2016spin}, promising a variety of interesting states. However, proving such a QSL ground state experimentally is a challenging task~\cite{savary2016quantum,wen2019experimental}. While the absence of long-range order (LRO) down to sub-Kelvin temperatures often hints at QSL behavior, alternative scenarios—such as disorder~\cite{zhu2017disorder}, spin glass (SG)~\cite{norman2016colloquium}, random singlet phases~\cite{song2021evidence}, eminuscent phases~\cite{syzranov2022eminuscent}, or ultra-low-temperature ordering~\cite{tokiwa2021frustrated}—complicate definitive identification. Thus, a comprehensive set of bulk measurements is essential to disentangle these possibilities and establish the true nature of the ground state.

\begin{figure*}[htb]
  \includegraphics[scale=0.52]{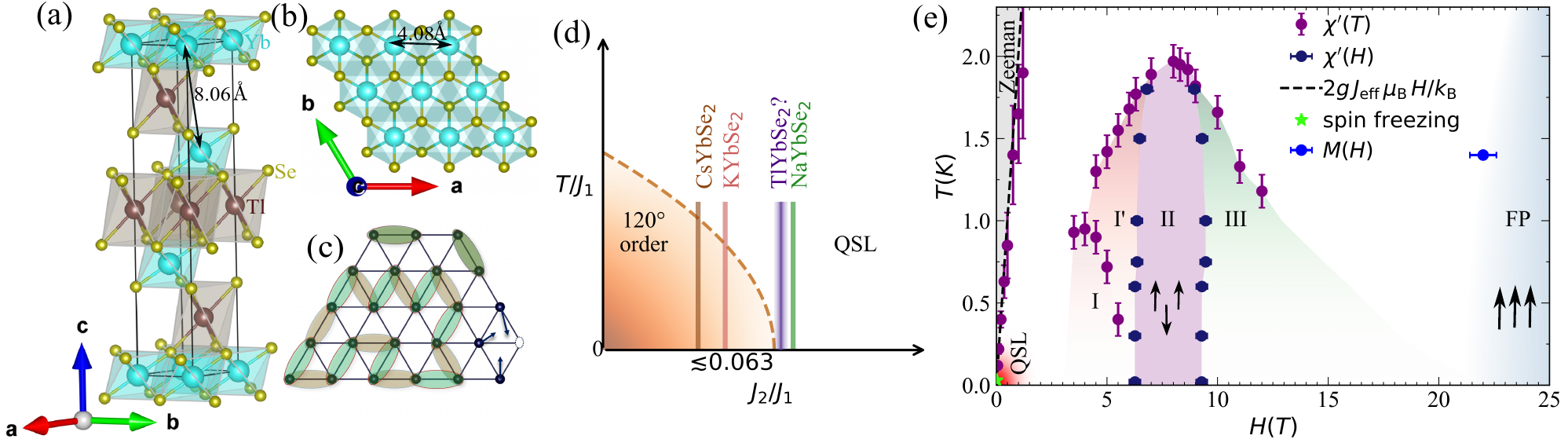}
\caption{Crystal structure and phase diagram.
(a) Crystal structure of TlYbSe$_2$, showing the stacking of Yb triangular layers separated by Tl layers along the $c-$axis. 
(b) Top view of the Yb triangular lattice in the $ab-$plane, where magnetic interactions are mediated by edge-sharing YbSe$_6$ octahedra. (c) A schematic representation highlighting the coexistence of a resonating valence bond (RVB) state with localized spin freezing in a triangular lattice geometry.
(d) Phase diagram showing the position of TlYbSe$_2$ in comparison to other triangular-lattice compounds in the $A$YbSe$_2$ family. The phase diagram and position of other compounds are adapted from Ref. \cite{scheie2024nonlinear,scheie2024spectrum}. The quantum critical point (QCP) at $J_2/J_1 \lesssim 0.063$ separates the $120^\circ$ ordered state and the quantum spin liquid (QSL) regime. The position of TlYbSe$_2$ is estimated from our experimental observations and lies near the QCP.  
(e) Temperature–field ($H$–$T$) phase diagram of TlYbSe$_2$ based on anomalies in the real part of the AC susceptibility $\chi'(T)$ (purple) and $\chi'(H)$ (dark blue). The dashed line indicates Zeeman-driven quenching of free Yb moments. The green star denotes a low-temperature spin freezing anomaly. A possible QSL phase is highlighted in red at low fields and temperatures. Field-induced magnetic phases are labeled I, I$'$, II, and III, with FP denoting the field-polarized state.
}
\label{structure}
\end{figure*} 

Recently, the rare-earth triangular lattice delafossite compounds have gained  attention~\cite{li2019ybmggao4,wen2019experimental,paddison2017continuous,scheie2024proximate,ranjith2019anisotropic,dai2021spinon,bordelon2019field,ding2019gapless,baenitz2018naybs,bag2024evidence,khatua2024magnetic,guo2019magnetism,ferreira2020frustrated}. Especially, the Yb$^{3+}$-based delafossites ($A$Yb$X_2$, with $A$ = monovalent cation; $X$ = O, S, Se, or Te), offer an ideal 2D triangular lattice geometry of effective spin-$1/2$ state of Yb$^{3+}$ ions. These systems offer a tunable interlayer and intralayer interactions via $A$- or $X$-site substitution~\cite{xing2021synthesis,xie2024rare}. Recently, some candidates of the family $A$Yb$X_2$ have already been shown to exhibit proximate QSL phases \cite{scheie2024proximate,bordelon2019field,ranjith2019anisotropic,dai2021spinon,zhang2021crystalline,zhang2022low,li2024thermodynamics,xing2019field,ding2019gapless,ranjith2019field,bordelon2020spin,baenitz2018naybs,sarkar2019quantum,xie2021field}. 
Interestingly,  NaYbSe$_2$ and KYbSe$_2$ have exhibited signatures of proximity to the quantum critical point (QCP) separating the LRO from QSL phases as fine-tuning the $A$-site cation enables control over the $J_2/J_1$ ratio~\cite{scheie2024nonlinear,scheie2024spectrum}. 

In this work, we introduce TlYbSe$_2$, a previously unexplored delafossite where thallium, a non-alkali metal, occupies the $A$-site. We report the low-temperature thermodynamic and magnetic properties of TlYbSe$_2$ using DC/AC magnetisation and heat capacity measurements down to $20$~mK. We find no evidence of magnetic LRO down to the lowest temperature. Both the heat capacity studies and the lattice constants place this compound between NaYbSe$_2$ and KYbSe$_2$ in the $J_2/J_1$ phase diagram [Fig.~\ref{structure}(d)] near the QCP leading to the QSL phase, where quantum fluctuations are strongest. We observe a frequency-dependent spin-freezing anomaly near $30$~mK—possibly originating from orphan spins—an effect not previously explored in any member of this family. Importantly, the magnetic heat capacity exhibits a robust linear-in-$T$ dependence below $350$~mK. 

The {\it linear} dependence of the heat capacity, also observed in isostructural NaYbSe$_2$~\cite{ranjith2019anisotropic}, is a conundrum because the theoretically expected QSL ground state for the $J_1$–$J_2$ Heisenberg model on the triangular lattice is a critical Dirac QSL ~\cite{alicea2005algebraic,iqbal2016spin,song2019unifying,willsher2025dynamics}, which is characterized by a {\it quadratic} temperature dependence of the specific heat. The latter arises from a linearly vanishing density of states (DOS), i.e. most simply arising from the Dirac spectrum of fermionic spinons in a fixed gauge flux background. Here we reconcile the observed heat capacity with a critical DSL by developing a phenomenological theory in which the thermal proliferation of gauge-flux excitations leads to a non-vanishing fermionic spinon density of states.

{\it Compound and thermodynamic measurements.--} 
TlYbSe$_2$ crystallizes in a delafossite-type structure with a trigonal $R\Bar{3}m$ (No.~166) space group. The structure consists of 2D triangular layers of edge-sharing, distorted YbSe$_6$ octahedra in the $ab-$plane, separated by TlSe$_6$ layers [Fig.~\ref{structure}(a,b)]. The Yb--Yb distance in a triangular layer is $\sim 4.08$~\AA. The unit cell contains three YbSe$_6$ layers stacked along the $c$-axis, with an interlayer Yb--Yb distance of $\sim 8.06$~\AA. Due to the larger ionic radius of Tl$^+$, the interlayer Yb--Yb spacing in TlYbSe$_2$ is notably larger than in most well-studied delafossites—such as NaYbSe$_2$ ($\sim 7.3$~\AA), NaYbO$_2$ ($\sim 5.85$~\AA), and KYbSe$_2$ ($\sim 7.93$~\AA)~\cite{bordelon2019field, ranjith2019anisotropic, scheie2024proximate}. This increased spacing is expected to weaken interlayer magnetic coupling, enhance two-dimensionality, and strengthen quantum fluctuations—conditions that may favor the emergence of a QSL ground state.

\begin{figure*}[!htb]
  \includegraphics[scale=0.5]{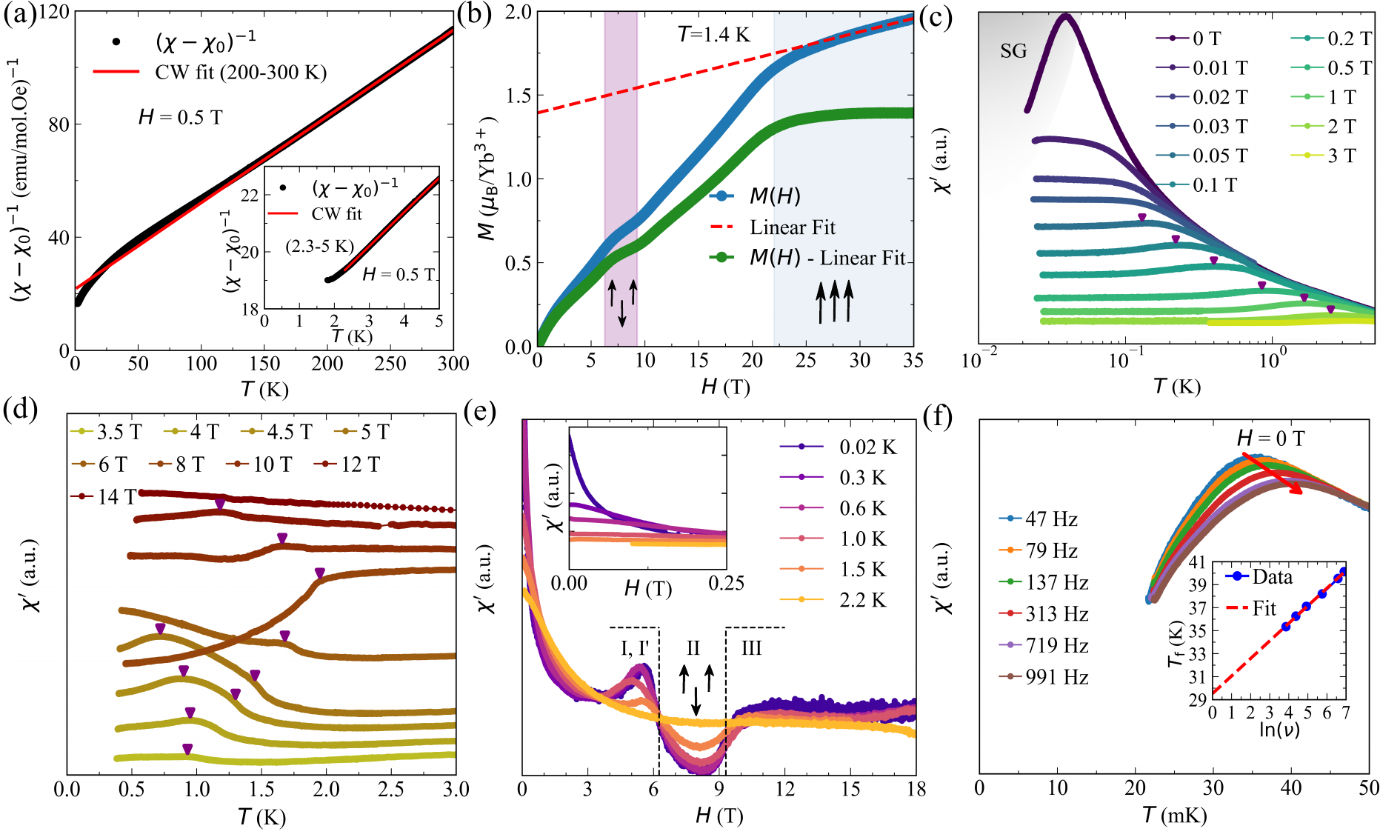}
\caption{AC and DC magnetization.
(a) Temperature-dependent inverse DC magnetic susceptibility ($1/\chi(T)$) at $0.5$~T, with a modified CW fit in the high-$T$ region. Inset: low-$T$ CW fit. 
(b) Isothermal magnetization $M(H)$ measured at $1.4$~K, showing saturation near $H_{\rm sat} \simeq 22$~T. The red dashed line represents a linear fit in the high-field region, yielding $M_{\rm sat} \simeq 1.4~\mu_{\rm B}$ per Yb$^{3+}$. The green curve represents the magnetization after subtracting the linear high-field slope from the raw $M(H)$ data. A magnetization plateau near $M_{\rm sat}/3$, characteristic of the up-up-down configuration in triangular-lattice antiferromagnets, is clearly visible. 
(c) Real part of AC susceptibility $\chi'(T)$ at low fields ($0$–$3$~T); purple arrows mark broad maxima due to Zeeman splitting. The light gray shaded region marks the spin-glass (SG) regime.
(d) $\chi'(T)$ at high fields ($3.5$–$14$~T), where the system
is in a magnetically ordered state. Curves are vertically offset for clarity (non-uniformly); original plots with no vertical offset data are provided in SM~\cite{supp}. Purple arrows mark transition temperatures. 
(e) Field-dependent $\chi'(H)$ data measured at various temperatures, with phase boundaries denoted by dotted black lines. The inset highlights the low-field behavior, showing a rapid suppression of $\chi'$ with increasing temperature.
(f) Frequency-dependent $\chi'(T)$ measured from $20$–$50$~mK at zero field, indicating a spin-glass transition likely arising from free spins ($y$-axis starts at origin). The red arrow marks the shift in the freezing temperature ($T_{\rm f}$) with frequency ($\nu$). Inset: $T_{\rm f}$ vs. $\ln(\nu)$, with a linear fit shown by the red dashed line.
}
\label{AC}
\end{figure*}

To investigate the nature and strength of the magnetic exchange interactions, we performed DC magnetization measurements [see Appendix~B]. The inverse magnetic susceptibility, $1/\chi(T)$, measured under an applied magnetic field of $0.5$~T [Fig.~\ref{AC}(a)], shows a linear behavior above $150$~K which can be described by the modified Curie-Weiss (CW) law [see supplemental Material (SM)~\cite{supp}]. The deviation from linear behavior below $100$~K suggests thermal depopulation of excited CEF levels. A low-$T$ CW fit below $5$~K yields $T$-independent susceptibility $\chi_0 \simeq 6.9 \times 10^{-3}$~emu/mol$\cdot$Oe, effective moment $\mu_{\rm eff}^{\rm LT} \simeq 2.6~\mu_{\rm B}$, CW temperature $\theta_{\rm CW}^{\rm LT} \simeq -13.8$~K, and a Land\'e $g$-factor $g \simeq 3.0$, consistent with a Kramers doublet $J_{\rm eff} = 1/2$ ground state. The large negative $\theta_{\rm CW}^{\rm LT}$ indicates dominant antiferromagnetic (AFM) exchange, with the nearest-neighbor (NN) exchange coupling estimated as $J_1 \simeq 9.2$~K using $\theta_{\rm CW} = -3J_1/2$. Notably, the values of $\theta_{\rm CW}$ and $J_1$ for TlYbSe$_2$ are much higher than those of its isostructural counterparts~\cite{ranjith2019field, ding2019gapless, bordelon2019field, ranjith2019anisotropic}, indicating stronger AFM exchange interactions.

The isothermal magnetization $M(H)$ measured at $1.4$~K [Fig.~\ref{AC}(b)] saturates near $H_{\rm sat} \simeq 22$~T, indicating strong AFM coupling. A finite slope beyond saturation reflects Van Vleck contributions, typical for Yb$^{3+}$ systems. A linear fit in the high-field range ($32$--$35$~T) yields a saturation moment $M_{\rm sat} \simeq 1.4~\mu_{\rm B}$, consistent with a $J_{\rm eff} = 1/2$ ground state and $g \simeq 2.8$. Using the relation $\mu_0 H_{\rm sat} = 9J_{\rm eff}k_{\rm B}/\mu_{\rm B}g$, the exchange coupling is estimated as $J_1 \simeq 9.2$~K, in excellent agreement with the value from the low-$T$ CW analysis.
The plateau near $M_{\rm sat}/3$ corresponds to the up-up-down (UUD) spin configuration—a hallmark of triangular-lattice antiferromagnets (TLAFs)~\cite{balents2010spin,chen2013ground,honecker2004magnetization,syromyatnikov2023unusual}.

To further understand the low-$T$ magnetism and field effects in TlYbSe$_2$, we performed AC susceptibility measurements [see Appendix~B]. The real part, $\chi'(T)$, measured at $137$~Hz under zero field, exhibits a sharp peak around $\sim 37$~mK [Fig.~\ref{AC}(c)], which shifts to higher temperatures with increasing frequency ($\nu$) [Fig.~\ref{AC}(f)]—a hallmark of spin freezing. A linear fit of $\ln(\nu)$ vs. freezing temperature ($T_{\rm f}$), extrapolated to $1$~Hz, yields a spin freezing transition temperature $T_{\rm g}=29.5(2)$~mK [Fig.~\ref{AC}(f) inset]. Applying a small magnetic field of $\sim0.02$~T suppresses the zero-field peak in $\chi'(T)$, consistent with the energy (temperature) scale of the spin freezing transition.
 
Upon further increasing field, a broad maximum appears [Fig.~\ref{AC}(c)] that shifts linearly, consistent with Zeeman splitting $\Delta E = 2\mu_{\rm B}gJ_{\rm eff}H$ of isolated Yb$^{3+}$ ($J_{\rm eff} = 1/2$) moments, such as those at grain boundaries of the polycrystalline sample. This aligns with the $M(H)$ data at $1.8$~K, indicating that $\sim 3$\% of the moments remain as free spins (see SM~\cite{supp}). A similar behavior was reported in NaYbO$_2$ with $\sim 7$\% free spins~\cite{bordelon2019field}.
With increasing magnetic field beyond $3.5$~T, TlYbSe$_2$ enters a magnetically ordered state [Fig.~\ref{AC}(d)]. Since AC susceptibility is less sensitive at high fields, anomalies in $\chi'(T)$ above $12$~T lie beyond the resolution. The $\chi'(H)$ curves at different temperatures [Fig.~\ref{AC}(e)] cross near $\sim 3.5$~T, where the system enters into an ordered state, followed by a peak at $\sim 5.5$~T and a dip at $\sim 8$~T—corresponding to the $M_{\rm sat}/3$ plateau in $M(H)$ [Fig.~\ref{AC}(b)]. Beyond $8$~T, $\chi'(H)$ recovers, marking a high-field boundary. 

\begin{figure*}[!htb]
  \includegraphics[scale=0.5]{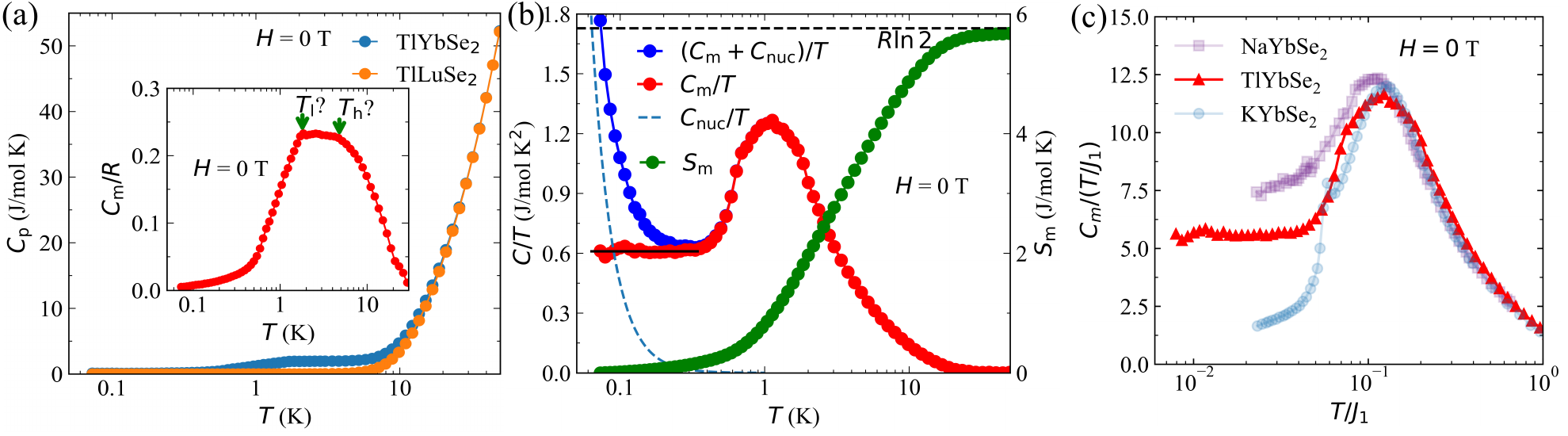}
\caption{Specific heat capacity. (a) Zero-field specific heat, $C_{\rm p}(T)$, of TlYbSe$_2$ measured down to 70~mK, overlaid with the phonon contribution $C_{\rm ph}(T)$ from non-magnetic TlLuSe$_2$. Inset: Magnetic heat capacity $C_{\rm m}/R$ showing two broad maxima near $T_{\rm l} \simeq 2$~K and $T_{\rm h} \simeq 5$~K.
(b) Zero-field heat capacity after subtracting $C_{\rm ph}(T)$: $C/T$ vs. $T$ (left $y$-axis) and magnetic entropy $S_{\rm m}$ vs. $T$ (right $y$-axis). The solid black line on $C_{\rm m}/T$ serves as a guide to the eye and indicates low-$T$ linearity. The dotted line marks $R\ln 2$ (right $y$-axis).
(c) Comparison of the magnetic specific heat $C_{\rm m}$ of TlYbSe$_2$ with KYbSe$_2$ and NaYbSe$_2$, plotted as $C_{\rm m}/(T/J_1)$ vs. $(T/J_1)$. The data for KYbSe$_2$ and NaYbSe$_2$ are adapted from Ref.~\cite{scheie2024spectrum}. TlYbSe$_2$ lies between the two at low $T$ and shows a more pronounced linear component.
}
\label{HC}
\end{figure*}

The temperature-field ($H$-$T$) phase diagram constructed from $\chi'(T)$, $\chi'(H)$, and $M(H)$ measurements is shown in Fig.~\ref{structure}(e). As the magnetic field increases, the system enters a field-induced magnetic LRO regime comprising multiple distinct phases. At low fields, the system transitions through phases~I and I$'$, followed by phase~II, centered around $6.3$--$9.3$~T, and then into phase~III, which extends up to the saturation field of $\sim22$~T. Phases~I and I$'$ likely correspond to $120^\circ$ or oblique (Y-coplanar) spin structures, and phase~II is attributed to the up-up-down (UUD) configuration, as evidenced by the suppression of $\chi'(H)$ [Fig.~\ref{AC}(e)] and the magnetization plateau at $M_{\rm sat}/3$ [Fig.~\ref{AC}(b)]. Phase~III has been proposed to correspond to a $2\!:\!1$ canted phase~\cite{seabra2011phase}.

To investigate the nature of the quantum disordered ground state in TlYbSe$_2$, we measured the zero-field specific heat $C_{\rm p}(T)$ down to $70$~mK [Fig.~\ref{HC}(a)], along with that of the nonmagnetic analog TlLuSe$_2$ to estimate the phonon background $C_{\rm ph}(T)$. The magnetic contribution, $C_{\rm m}(T)$, was obtained by subtracting $C_{\rm ph}(T)$ and the low-$T$ nuclear Schottky term $C_{\rm nuc}(T)$, extracted via fitting (discussed below). The $C_{\rm m}(T)$ shows no sign of magnetic LRO, consistent with AC susceptibility results. Instead, a broad maximum is observed between $2-5$~K [inset, Fig.~\ref{HC}(a)], characteristic of low-dimensional frustrated systems and indicative of a crossover into a QSL regime.

Notably, related compounds such as NaYbO$_2$ and NaYbSe$_2$ exhibit similar broad maxima arising from a superposition of two broad peaks, suggesting two distinct exchange energy scales~\cite{ranjith2019anisotropic,bordelon2019field}. Such double maxima in $C_{\rm p}(T)$ have been theoretically predicted for 2D triangular and kagome antiferromagnets with fully frustrated disordered ground states~\cite{ishida1997low,elser1989nuclear}, and observed experimentally in several TLAF systems~\cite{ranjith2019anisotropic,bordelon2019field,rawl2017ba,cui2018mermin}. For spin-$1/2$ TLAFs, the two crossover temperatures are expected at $T_{\rm l}/J = 0.2$ and $T_{\rm h}/J = 0.55$~\cite{chen2019two}. With $J \simeq 9.2$~K, the observed maxima in TlYbSe$_2$ fall near these theoretical values. The peak value $C_{\rm m}^{\rm max} \simeq 0.23R$ also agrees with predictions for fully frustrated TLAFs~\cite{elstner1993finite,bernu2001specific}. 

The magnetic entropy $S_{\rm m}$, obtained by integrating $C_{\rm m}/T$ from $70$~mK, recovers $\sim 98\%$ of $R\ln2$ by $20$~K [Fig.~\ref{HC}(b)], confirming a Kramers doublet ground state with effective spin-$1/2$ for Yb$^{3+}$. The near-complete entropy recovery implies minimal residual entropy below $70$~mK. If the spin freezing seen in $\chi'(T)$ were intrinsic, it would contribute significant low-$T$ entropy~\cite{Mydosh_2015}. The glassy behavior at zero field is therefore attributed to the $\sim 3\%$ orphan Yb$^{3+}$ moments identified in magnetization measurements. Selenium vacancies may also cause Yb spin freezing, implying the orphan spins could stem from such defects. Similar behavior has been reported in the disorder-free Dirac QSL candidate YbZn$_2$GaO$_5$~\cite{bag2024evidence}. We conclude that these orphan moments account for the observed spin freezing, while the rest of the lattice retains highly fluctuating spins down to the lowest temperatures.

Having established a dynamic ground state persisting down to $20$~mK, we now focus on the low-$T$ heat capacity to further probe its nature. The zero-field $C/T$ after subtracting the phonon part is shown in Fig.~\ref{HC}(b) (left $y$-axis). Below $300$~mK, an upturn appears due to a nuclear Schottky contribution—typical for Yb-based systems~\cite{steppke2010nuclear,ranjith2019anisotropic}. 
The low-$T$ data ($T < 350$~mK) were fitted using $C(T) = \alpha/T^2 + \gamma T^b$, where the first term captures the nuclear Schottky contribution ($C_{\rm nuc}$) and the second term reflects the intrinsic magnetic part ($C_{\rm m}$). The fit yields $\alpha = 4.468(75) \times 10^{-4}$~J\,K/mol, $\gamma = 0.62(1)$~J/mol\,K$^2$, and $b = 1.01(1)$. For a $U(1)$ Dirac spin liquid, the magnetic heat capacity is expected to follow a $T^2$ dependence. However, a pure $T^2$ fit fails to describe the observed behavior [see SM~\cite{supp}, Fig.~S4(c)]. 

Interestingly, the low-$T$ magnetic heat capacity of TlYbSe$_2$ lies between that of KYbSe$_2$ and NaYbSe$_2$~\cite{scheie2024spectrum}, as shown in Fig.~\ref{HC}(c), placing it between the two in the $J_2/J_1$ phase diagram [Fig.~\ref{structure}(d)]. Combined with the absence of $120^\circ$ magnetic order in the finite-field regime below $2$~T and down to $20$~mK, these findings suggest that TlYbSe$_2$ lies closer to the QCP than other delafossite candidates. This trend in heat capacity faithfully reflects the distances Yb - Yb within the triangular layer, which increase from 4.06~\AA\ in NaYbSe$_2$ to $4.08$~\AA\ in TlYbSe$_2$, $4.11$~\AA\ in KYbSe$_2$, and $4.15$~\AA\ in CsYbSe$_2$. This structural trend further supports the placement of TlYbSe$_2$ in the phase diagram and highlights Yb--Yb spacing as a useful indicator for bulk thermodynamic behavior and proximity to a QSL state. 
Notably, the key results presented here ---  such as the observation of spin freezing at ultra-low temperatures and linear-in-$T$ specific heat --- were obtained under zero field, meaning they are independent of the crystallographic field direction. As such, measurements on polycrystalline and single-crystalline samples are expected to yield similar results.  

{\it Thermal flux proliferation.--}
The near-linear $T$-dependence ($b \simeq 1$) of magnetic specific heat $C_{\rm m}(T)$, along with the absence of magnetic LRO down to $20$~mK, suggests a non-vanishing DOS at low energy in the QSL phase~\cite{zhou2017quantum}. 
Such linear-in-$T$ specific heat arises naturally in a QSL with a spinon Fermi surface but no microscopic models and mechanisms are known which stabilize a spinon Fermi surface state on the triangular lattice~\cite{iqbal2016spin}. In the following, we show how thermally excited fluxes above a U(1) Dirac QSL ground state can generate a \emph{finite} spinon DOS at low energy inducing a linear $C_{\rm m}(T)$. We note that our proposed mechanism is the U(1) analogue of the enhancement in low-energy DOS produced by fluctuating $Z_{2}$ fluxes in the Kitaev spin liquid~\cite{Nasu15,penghao24}. 

\begin{figure}[t]
    \centering
    \includegraphics[width=\linewidth]{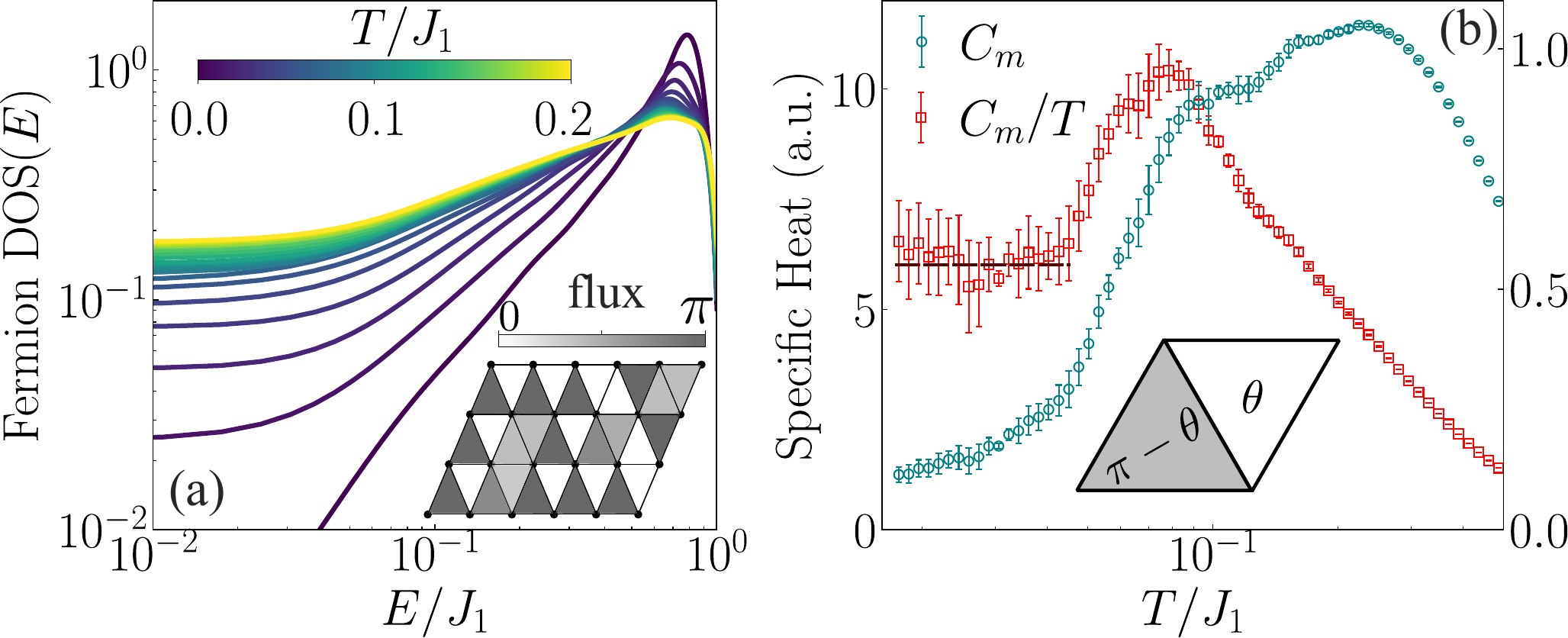}
    \caption{Results from the phenomenological theory. (a) Fermionic spinon density of states (DOS) as a function of temperature. The temperature sets the density of excited gauge fluxes, see inset for an  illustration of a flux configuration. (b) Specific heat presented as $C_m/T$ and $C_m$, with the dashed horizontal line as a guide to the eye for the linear specific heat.  The inset shows a local flux variation with phase angle $\theta$ on top of the staggered flux configurations of the ground state Dirac QSL. Data are obtained by importance sampling for the effective model described in the text. }
    \label{fig:doscv}
\end{figure}

Following the standard parton mean-field theory we assume the fermionic spinons propagate in a staggered $\pi$-flux background giving a Dirac spectrum. We assume a separation of scale such that the dynamics of the background gauge field excitations are decoupled from the spinons - the latter see a given gauge field configuration as a static background flux. This can be described by an effective Hamiltonian $H = \sum_i E_M^i(F) \,a_i^\dagger a_i +  \sum_{\theta \in [0, 2\pi)}\sum_m \Delta (\theta) (f^\theta_m)^\dagger f^\theta_m$, where $a_i^{\dagger}$ create matter fermions with energy  $E^{i}_{M}(F)$ in a given flux background $F$, and $(f^\theta_m)^\dagger$ create a local U(1) flux excitation with a phase angle~$\theta$ and gap $\Delta(\theta)$, see insets of Fig.\ref{fig:doscv},(a,b). Note, because of the U(1) nature of the gauge field local flux excitations are gapless in contrast to the Kitaev $Z_2$ case. We can then perform importance sampling of the joint flux–fermion distribution. We find that the fermion DOS, which vanishes linear in energy at zero temperature without local flux excitations, becomes finite once thermal fluxes proliferate under finite $T$ [Fig.\ref{fig:doscv},(a)].  Immediately above the temperature scale $T/J_1\sim 10^{-2}$ at which this DOS becomes a finite constant, the specific heat capacity follows $C_m\propto T$ [Fig.~\ref{fig:doscv},(b)], and even quantitatively matches the linear specific heat observed experimentally in Fig.~\ref{HC}.  Thus, the finite-$T$ thermodynamics of a zero-temperature U(1) Dirac spin liquid can be fully consistent with our data; Details of the phenomenological model are given in the End-Matter and the Supplementary Material \cite{supp}. 

{\it Conclusion.--}
The powder results presented here are dominated by grains aligned with the $ab-$plane - statistically responsible for roughly 2/3rd of the contribution. Also, a strong magnetic anisotropy in delafossites typically causes features associated with $H||c$ to appear only at very high fields, likely beyond the field range accessed in this study~\cite{ranjith2019anisotropic}. As a result, even though our measurements were performed on polycrystalline samples with randomly oriented grains, the field-induced transitions remain sharp--comparable to those seen in single crystals. Consequently, the use of polycrystalline samples does not compromise the accuracy of the phase diagram, which closely resembles those reported for related delafossites under $ab-$plane magnetic fields~\cite{lee2024magnetic, ranjith2019anisotropic}.
Importantly, the linear-in-$T$  trend of the low-temperature magnetic heat capacity is observed in zero-field measurement, making it independent of field orientation and therefore unaffected by whether the sample is single-crystalline or polycrystalline. 
We have shown that this behavior is consistent with a non-vanishing spinon DOS and is well described by a phenomenological model of a gapless Dirac $U(1)$ spin liquid with thermally excited gauge fluxes. Comparative analysis of heat capacity data places TlYbSe$_2$ between NaYbSe$_2$ and KYbSe$_2$ in the $J_2/J_1$ phase diagram, suggesting closer proximity to a QCP separating magnetic order from a spin-liquid regime. The observation of spin-freezing transition at ultra-low temperatures, likely originating from orphan spins, calls for future studies of similar low-temperature spin-freezing in other delafossites, and cautions against readily interpreting low-temperature anomalies as spin gaps.

Altogether, our results establish TlYbSe$_2$ as a promising platform for studying QSL physics and field-induced magnetism in triangular-lattice systems. Future studies on single crystals—including neutron scattering and thermal transport—will be essential for probing potential fractionalized excitations and the underlying spin dynamics. It will also be important to explore theoretically whether the thermal proliferation of gauge fluxes, which account for the observed specific heat, could also be responsible for the spin freezing and alter the dynamical response.

\acknowledgments 
\noindent We thank A. O. Scheie, D. D. Sarma, and J. Xing for the fruitful discussions and valuable suggestions. B.P.B., A.B., and the research as a whole is supported by the U.S. Department of Energy – Office of Science, Basic Energy Sciences (Grant No. DE-SC0022986), under the project "\textit{Seeking Quasiparticles from Low-Energy Spin Dynamics.}" A.U. acknowledges financial support from the Department of Science and Technology (DST), Government of India, through the DST Inspire Faculty Fellowship (Ref. No. DST/INSPIRE/04/2019/001664), as well as from the Quantum Science Center, a National Quantum Initiative Science Research Center funded by the U.S. Department of Energy. 
S.F. and J.K. acknowledge support from  Deutsche Forschungsgemeinschaft (DFG, German Research Foundation) under Germany’s Excellence Strategy.--EXC.--2111.--3908141008 as well as the Munich Quantum Valley, which is supported by the Bavarian state government with funds from the Hightech Agenda Bayern Plus. J.K. also acknowledges support from the Imperial-TUM flagship partnership.
Part of this work was conducted at the National High Magnetic Field Laboratory, supported by the National Science Foundation (Cooperative Agreement No. DMR-2128556) and the State of Florida. B.P.B. acknowledges the Quantum Design Winter School, where milli-Kelvin specific heat capacity data were collected.\\

\bibliographystyle{apsrev}
\bibliography{references_v1}

\onecolumngrid
\vspace{1em}
\begin{center}
    \textbf{End Matter}
\end{center}
\vspace{1em}
\twocolumngrid

\textit{Appendix A: Sample Preparation $-$}Polycrystalline samples of TlYbSe$_2$  were synthesized via the solid-state reaction method using thallium (Tl) granules (99.99\%, Thermo Fisher), ytterbium (Yb) powder (99.9\%, Thermo Fisher), and selenium (Se) powder (99.999\%, Thermo Fisher) mixed in appropriate ratios and maintained at $800~^\circ$C for $72$ hours yielding a black powder. The non-magnetic analog TlLuSe$_2$, used as a reference for phonon heat capacity, was synthesized following an identical procedure using lutetium (Lu) powder (99.9\%, Thermo Fisher) instead of Yb powder. The phase purity of the samples was confirmed by performing a Le Bail profile fit on powder XRD patterns recorded at room temperature using a PANalytical Empyrean diffractometer equipped with an incident-beam monochromator and Cu K$_{\alpha}$ radiation.\\

\textit{Appendix B: Magnetic measurements $-$} DC magnetization measurements were performed down to 1.8~K and up to 7~T using a Quantum Design SQUID magnetometer (MPMS-3) at the Birck Nanotechnology Center user facility, Purdue University. High-field isothermal magnetization up to 35~T was measured using a vibrating sample magnetometer (VSM) at the National High Magnetic Field Laboratory (NHMFL).

AC magnetization measurements were conducted at NHMFL using two different superconducting magnet systems. Measurements down to $300$~mK were carried out using the $18$~T General Purpose Superconducting Magnet (SCM2) equipped with a $^3$He insert. These measurements were performed in an AC susceptometer at a frequency of $313$~Hz with an AC field of $1.7$~Oe. To access lower temperatures, measurements between $20$~mK and $1$~K were performed using the $18$~T superconducting magnet with a top-loading dilution refrigerator (SCM1). In this setup, an AC field of $0.9$~Oe at $137$~Hz was applied.\\

\textit{Appendix C: Heat capacity measurements $-$} Specific heat capacity measurements were carried out using a Physical Property Measurement System (PPMS, Quantum Design). Field-dependent measurements down to $1.8$~K were performed at the Spin Lab facility in the Birck Nanotechnology Center, Purdue University. Zero-field measurements extending down to $70$~mK were conducted using a dilution refrigerator insert at Quantum Design’s headquarters in San Diego, California. For measurements below $1.8$~K, the sample was mixed with silver in a $1:1$ ratio to improve thermal conductivity and then pressed into a pellet. The heat capacity of the sample was obtained by subtracting the known contribution from silver.\\

\textit{Appendix D: Phenomenological theory for the linear specific heat $-$} 
We present a simple phenomenological model explaining the observed linear specific heat. 
For a concrete demonstration, we start from the effective parton mean-field Hamiltonian~\cite{iqbal2016spin,willsher2025dynamics}: $H \sim \sum_{\expval{ij}} \sum_\alpha \chi_{ij} c_{i,\alpha}^\dagger c_{j,\alpha} + \rm const.$,
with $\chi_{ij}$ a mean-field parameter in a staggered $\pi$-flux background. 
We assume a bilinear Hamiltonian treating the coupling of  fermionic spinons to gauge field excitations as a classical background field similar to the conserved $Z_2$ gauge field in Kitaev QSLs~\cite{Nasu15}:
\begin{equation} \label{eq:hamf}
    H = \sum_i E_M^i(F) \,a_i^\dagger a_i +  \sum_{\theta \in [0, 2\pi)}\sum_m \Delta (\theta) (f^\theta_m)^\dagger f^\theta_m
\end{equation}
where $E_M^i(F)$ is the $i$-th single spinon energy of the matter fermions conditioned on the flux sector. The latter is determined by the occupation of local flux 'fermions' $f_m^\theta$ with $\Delta (\theta)$ the energy of the local U(1) flux excitation. 
Note that we have essentially decoupled the spinon dynamics from the gauge field, which is at odds with a strong coupling low-energy theory of the U(1) DSL in terms of emergent QED$_{2+1}$~\cite{song2019unifying}. Therefore, we expect it to fail for asymptotically low temperatures. Nevertheless, we expect it to capture the essence of the low but {\it finite} temperature regime where the dynamics of flux excitations is much slower than the fast spinons. 
We can, thus, use our Hamiltonian to derive  a simple partition function
\begin{equation} \label{eq:Z}
    Z = {\rm Tr}_{F,M} e^{-\beta H} =\sum_{F=\{n^{\theta}_m\}} e^{-\beta\sum_{\theta;m} \Delta(\theta)\,n^\theta_m}\, Z_m[F]
\end{equation}
where $n^\theta_m =(f_m^\theta)^\dagger f_m^\theta$ is the occupation number of flux deviating from the ground state by $\theta$, $M_F$ denotes states in the matter fermion sector conditioned on the flux configuration $F$, and $Z_m[F]$ is the fermion partition function conditioned on $F$. 
Since the fermion and flux are correlated, the probability distribution of fluxes is determined by the effective total flux energy $\mathcal{F}_{\rm eff}(F)$
\begin{equation}
    W(F)=\frac{e^{-\beta\mathcal{F}_{\rm eff}(F)}}{Z},\;
    \mathcal{F}_{\rm eff}(F)
    = \sum_{\theta,m}\Delta(\theta) n_m^\theta - T \ln Z_m[F],
\end{equation}
Using the law of total variance, the specific heat $C_V(\beta) = \beta^2 \left[\expval{E^2} - \expval{E}^2 \right]$ can be computed by
\begin{align}
    C_V(\beta)
    = \beta^2 \,\expval{{\rm Var}(E)}_F + \beta^2 \,{\rm Var}_F(\expval{E}) \label{eq:Cv}
\end{align}
with $\expval{{\rm Var}(E)}_F$ the \emph{averaged variance} of total energy $E$ (conditioned on a $F$) over different flux sectors; and ${\rm Var}_F(\expval{E})$ the \emph{variance of the expectation of total energy} across different flux sectors $F$. 

Figure~\ref{fig:doscv}(b) shows that $C_m/T$ is nearly constant at temperatures $T/J_1 \sim 10^{-2}$, then rises to a broad maximum, precisely the behavior reported experimentally.
This regime is reached once flux excitations with all phase angles $\theta\in[0,2\pi)$ become thermally populated. The resulting spinon band structure develops a $\emph{finite}$ density of states near the Fermi level, and  even a modest flux density produces a finite ${\rm DOS}(E \to 0)$.  A standard low-$T$ expansion with this finite ${\rm DOS}(E\to 0)$ then yields the observed linear specific-heat coefficient $C_m/T\sim \rm const.$.  
We point out that at even lower temperatures a Schottky-like upturn in the specific heat could emerge from the two-level $n^\theta$ \cite{supp} -- an effect also observed in $Z_{2}$ spin liquids \cite{Nasu15} and NaYbSe$_2$~\cite{scheie2024spectrum}.  This low-$T$ anomaly, however, is irrelevant for the fermionic thermodynamics discussed here: the supplemental analysis shows that $\Delta(\theta)$ is exceedingly small, hence throughout the temperature regime where $k_BT$ is larger than the effective flux gap, the majority contribution to the magnetic specific heat comes from fermionic spinons. More detailed discussions can be found in \cite{supp}.

\end{document}